# Electron transport through tripeptides self-assembled monolayers

*Evgeniy Mervinetsky,[1,2] Israel Alshanski,[1,2] Stephane Lenfant,[3] David Guerin,[3]*

*Leonardo Medrano Sandonas,[4,6] Arezoo Dianat,[4] Rafael Gutierrez,[4]*

*Gianaurelio Cuniberti,[4,5,6] Mattan Hurevich,[1,2]*

*Shlomo Yitzchaik,[1,2] and Dominique Vuillaume[3]*

1) Institute of Chemistry, The Hebrew University of Jerusalem, Safra campus, Givat Ram, Jerusalem, 91904, Israel.

2) Center for Nanoscience and Nanotechnology, The Hebrew University of Jerusalem, Safra campus, Jerusalem, 91904, Israel.

3) Institute for Electronics Microelectronics and Nanotechnology, CNRS, Univ. Lille, 59652, Villeneuve d'Ascq, France.

4) Institute for Materials Science and Max Bergmann Center of Biomaterials, TU Dresden, 01069 Dresden, Germany.

5) Dresden Center for Computational Materials Science, TU Dresden, 01062 Dresden, Germany.

6) Center for Advancing Electronics Dresden, TU Dresden, 01062 Dresden, Germany

Emails : rafael.gutierrez@tu-dresden.de; shlomo.yitzchaik@mail.huji.ac.il; dominique.vuillaume@iemn.fr

ABSTRACT. We report how the electron transport through a solid-state metal/ Gly-Gly-His tripeptide (GGH) monolayer/metal junction and the metal/GGH work function are modified by the GGH complexation with $Cu^{2+}$ ions. Conducting AFM is used to measure the current-voltage histograms. The work function is characterized by combining macroscopic Kelvin probe and Kelvin probe force microscopy at the nanoscale. We observe that the $Cu^{2+}$ ions complexation with the GGH monolayer is highly dependent on the molecular surface density and results in opposite trends. In the case of a high density monolayer the conformational changes are hindered by the proximity of the neighboring peptides, hence forming an insulating layer in response to copper-complexation. Whereas the slightly lower density monolayers allow for the conformational change to a looped peptide wrapping the Cu-ion, which results in a more conductive monolayer. Copper-ion complexation to the high- and low-density monolayers systematically induces an increase of the work functions. Copper-ion complexation to the low-density monolayer induces an increase of electron transport efficiency, while the copper-ion complexation to the high-density monolayer results in a slight decrease of electron transport. Both of the observed trends are in agreement with first-principle calculations. Complexed copper to low density GGH-monolayer induces a new gap state slightly above the Au Fermi energy that is absent in the high density monolayer.

**INTRODUCTION**

Molecular devices made of peptides molecules and proteins are gaining interest as nanoscale devices in bioelectronics.[1-2] Understanding the electron transport mechanisms through these biomolecules is a key issue in biology. Similarly, biomolecules can be used to build and study various bioelectronic nanomaterials and devices.[3-6] For example, the "doping" of a polypeptide chain (7-alanine) by one single tryptophan substitution enhances the electron transport, a mechanism ascribed to the introduction of an additional energy level close to the Fermi energy of the Au electrodes.[4] Single molecule measurements (scanning tunneling microscope break junction) show that the conductance of a metal/polypeptide/metal junction can be controlled by the pH of the surrounding media due to a conformational change from a compact helical to a more extended structure,[7] and pH titration was also demonstrated from single peptide conductance measurements.[8] The electron transport through a helical peptide is also dependent whether the electrons are injected parallel or anti-parallel to the peptide dipole.[9] Combining peptides and redox species (e.g. ferrocene),[10-12] or π-conjugated moieties,[3] is also a powerful way to modify the electron transport properties of the peptides and tailor electronic functionality in biologically relevant macromolecules. Spin-dependent transport through chiral peptides has also been demonstrated as an example of the CISS (chirality induced spin selectivity) effect.[13]

Some of us have reported that a self-assembled monolayer of Gly-Gly-His tripeptide (GGH) is a very sensitive electrochemical sensor of copper ions.[14] The GGH molecules in the monolayer chelate with the $Cu^{2+}$ ions by a conformational change, forming a dense barrier, which prevent the access of redox species to the electrode during the electrochemical impedance spectroscopy measurements, leading to an increase of the electrochemical impedance.

Here, we study the electronic properties of a solid-state metal/GGH monolayer/metal junction and we examine how they are modified by the GGH reaction with $Cu^{2+}$ ions. The electron transport through the metal/GGH monolayer/metal junction is studied by measuring the current versus voltage curves histograms at the nanoscale with a conducting AFM. The work function of metal/GGH is characterized by combining macroscopic Kelvin probe and Kelvin probe force microscopy at the nanoscale. The self-assembled monolayers (SAM) of GGH were also characterized by ellipsometry, XPS and topographic AFM to assess the formation of the SAM on gold ultra-flat electrodes and the reaction with $Cu^{2+}$ ions. We study two samples prepared to have a slightly different molecular packing of the molecules in the SAMs. In both cases, we observe that the $Cu^{2+}$ ions reaction with the GGH monolayer systematically induces an increase of the work function, in agreement with Density-Functional Theory (DFT) calculations. The effect of $Cu^{2+}$ exposure on the electron transport through the metal/GGH/C-AFM junction depends on the GGH molecular packing. For the denser SAM, the current passing though the metal/GGH/C-AFM junctions is slightly reduced when exposed to $Cu^{2+}$ ions, while it is increased (about a decade) for the slightly less dense SAM. DFT calculations help to rationalize these results. In the less dense case, the GGH undergoes a large conformational change (GGH folding around the $Cu^{2+}$) to fully chelate the $Cu^{2+}$ ion ($Cu^{2+}$ surrounded by the 4 N atoms of the GGH) with square-planar configuration, while in the denser SAM, due to steric hindrance, the $Cu^{2+}$ ions are partly chelated with less than 4 N atoms of the GGH (figure 1). In this former case, DFT calculations show that the increase of electron transport is consistent with the modification of the molecular orbitals at the metal/GGH system and the appearance of metal (Cu) induced gap state slightly above the Au Fermi energy.

**METHODS**

*GGH self-assembled monolayers* - The SAMs were formed on template-stripped Au ($^{TS}$Au) electrodes. The very flat $^{TS}$Au surfaces were prepared according to the method developed by the Whitesides group.[15] In brief, a 300–500 nm thick Au film was evaporated on a very flat silicon wafer covered by its native $SiO_2$ (rms roughness of ~0.4 nm), which was previously carefully cleaned by piranha solution (30 min in 7:3 $H_2SO_4$/$H_2O_2$ (v/v); Caution: Piranha solution is a strong oxidizer and reacts exothermically with organics), rinsed with deionized (DI) water, and dried under a stream of nitrogen. A clean glass piece (ultrasonicated in acetone for 5 min, ultrasonicated in 2-propanol for 5 min, and UV irradiated in ozone for 10 min) was glued (UV polymerizable glue) on the evaporated Au film and mechanically stripped with the Au film attached on the glass piece (Au film is cut with a razor blade around the glass piece). This very flat (rms roughness of ~0.4 nm, the same as the $SiO_2$ surface used as template) and clean template-stripped $^{TS}$Au surface was immediately used for the formation of the SAM.

*Ellipsometry* - We recorded spectroscopic ellipsometry data in the visible range using an UVISEL (Jobin Yvon Horiba) spectroscopic ellipsometer equipped with DeltaPsi 2 data analysis software. The system acquired a spectrum ranging from 2 to 4.5 eV (corresponding to 300 to 750 nm) with intervals of 0.1 eV (or 15 nm). Data were taken at an angle of incidence of 70°, and the compensator was set at 45°. Data were fitted by a regression analysis to a film-on-substrate model as described by their thickness and their complex refractive indexes. First, a background before monolayer deposition for the gold coated substrate was recorded. Secondly, after the monolayer deposition, we used a 2-layer model (substrate/SAM) to fit the measured data and to determine the SAM thickness. We employed the previously measured optical properties of the gold coated substrate (background), and we fixed the refractive index of the organic monolayer at 1.50. The usual values in the literature for the refractive index of organic monolayers are in the range 1.45–1.50.[16] We can notice that a change from 1.50 to 1.55 would result in less than 1 Å error for a thickness less than 30 Å. We estimated the accuracy of the SAM thickness measurements at ± 2 Å.

*X-ray Photoelectron Spectroscopy* - XPS was performed with a Physical Electronics 5600 spectrometer fitted in an UHV chamber with a residual pressure of 2×10$^{-10}$ Torr. High resolution spectra were recorded with a monochromatic Al Kα X-ray source (hν = 1486.6 eV), a detection angle of 45° as referenced to the sample surface, an analyzer entrance slit width of 400 $\mu$m and with an analyzer pass energy of 12 eV. Semi-quantitative analysis were completed after standard background subtraction according to Shirley's method.[17] Peaks were decomposed by using Voigt functions and a least-square minimization procedure and by keeping constant the Gaussian and Lorentzian broadenings for each component of a given peak.

*Kelvin Probe* - Contact potential difference (CPD) was measured on large area samples with Kelvin Probe S (DeltaPhi Besocke, Jülich, Germany), with a vibrating gold electrode (work function 5.1 eV) in a home-built Faraday cage under Ar (argon) atmosphere.

*Kelvin Probe Force Microscopy* - KPFM measurements were carried out at room temperature with a Dimension 3100 from Veeco Inc., purged with a flow of dry nitrogen atmosphere. We used Pt/Ir tip (PPP-EFM-50 from Nanosensors) with spring constant of ca. 3 N/m and a resonance frequency of ca. 70 kHz. Topography (tapping mode AFM) and KPFM data were recorded using a standard two-pass procedure, in which each topography line acquired in tapping mode is followed by the acquisition of KPFM data in a lift mode, with the tip scanned at a distance z ~ 100 nm above the sample so as to discard short range surface forces and be only sensitive to electrostatic forces. DC and AC biases ($V_{DC} + V_{AC} \sin(\omega t)$) are applied to the cantilever with $V_{AC}$ = 2 V. Experimentally, the contact potential difference (CPD) is measured using a feedback loop which sets to zero the cantilever oscillation amplitude by adjusting the tip DC bias $V_{DC}$. The work function (WF) of the sample was deduced from the CPD following the relation, WF = $W_{tip}$ - e.CPD, where e is the elementary charge and $W_{tip}$ the work function of the KPFM tip. Since we focused on the WF modifications, the exact value of $W_{tip}$ is not an issue, and we simply have $\delta$WF = - e.$\delta$CPD. We recorded images (1 μm x 1 μm) at 3-4 different zones on the sample, from which we constructed the CPD histograms.

*Conducting Atomic Force Microscope* - Conducting atomic force microscopy (C-AFM) was performed under a flux of $N_2$ gas (ICON, Bruker), using a tip probe in platinum/iridium (SCM-PIC v2 from Bruker). The tip loading force on the surface was fixed ≤ 50 nN to avoid a too important strain-induced deformation of the monolayer (≤ 0.3 nm).[18] A square grid of 10×10 was defined with a pitch of 100 nm. At each point, the I-V curve is acquired leading to the measurements of 100 I-V traces. This process was repeated 3 times at different places on the sample, and the 300 I-V traces were used to construct the current-voltage histograms. The bias was applied on the $^{TS}$Au substrate and the tip was grounded through the input of the current amplifier. Note that around 0V the currents are very weak and in many cases at the limit of detection (0.1 pA). Thus it is likely that any "asymmetry" in the IV around 0V induced by the existing surface potential (see KPFM) in the SAMs is not observable.

*Computational modeling* - The adsorption of Lpa-GGH onto Au(111) surface, the corresponding interaction properties with $Cu^{2+}$ ions as well as the calculation of the work function were theoretically addressed at the Density-Functional Theory (DFT) level. We used mixed Gaussian plane wave (GPW) methods with the standard implementation in the CP2K package.[19] Here, the Kohn-Sham orbitals are expanded into linear combinations of contracted Gaussian type orbitals and complemented by a plane-wave basis set in order to compute the electronic charge density. In all calculations, the PBE (Perdew, Burke, and Ernzerhof) exchange-correlation functional was used,[20] and its corresponding norm-conserving pseudo-potential GTH (Goedecker, Teter and Hutter).[21] Finally, a DZVP (double zeta for valence electrons plus polarization functions) basis set complemented with a plane-wave basis set energy cut-off 350 Ry was employed and dispersion corrections were included through the standard D2-Grimme parameterization.[22]

**RESULTS**

**GGH self-assembled monolayers**

The synthesis of the Lpa-GGH (lipoic acid Gly-Gly-His) tripeptide was reported by some of us in a previous work.[14] Formation of low- and high-density tripeptide monolayers (LDP and HDP respectively) were prepared by dipping cleaned $^{TS}$Au substrates (see Methods) either in a 10 μM or 250 μM solution of Lpa-GGH in DI water for 2h, then rinsed copiously with DI water and dried under dry $N_2$ stream. The change in the concentration resulted in a LDP and HDP monolayers (see below).

For the $Cu^{2+}$ exposure, both LDP and HDP monolayers were dipped overnight into a solution of 10 μM $Cu(NO_3)_2$ and 640 μM $HNO_3$ in DI water, then rinsed copiously with DI water and dried under dry $N_2$ stream. To avoid any measurement bias and irreproducible results, we used two protocols (namely A and B, Fig. S1 in the Supporting Information) to study the effect of $Cu^{2+}$ exposure on electron transport and work function. We started by preparing two samples in the same Lpa-GGH solution. Sample A was characterized after the SAM formation and then exposed to $Cu^{2+}$ and measured again. As a control experiment, sample B was immediately dipped in the $Cu^{2+}$ solution and then measured. Protocol B checks that the measurements before $Cu^{2+}$ exposure have not induced any contamination during the measurements of the as-prepared sample. Protocol A ensures that the comparison without and with $Cu^{2+}$ was made on the same sample. Thus, if we observe the same trends (at least qualitatively) between samples A and B, we are reasonably sure that we have no measurement bias, no contamination and no strong irreproducibility. We mainly report data obtained with protocol A (unless specified), and the control data (protocol B) are given in the supporting information.

**Physical characterization of the GGH SAM**

First, the TSAu-Lpa-GGH SAMs were characterized by spectroscopic ellipsometry to determine their thicknesses. The high-density peptide (HDP) monolayers have a thickness of 2.2 ± 0.2 nm in good agreement with the calculated length (2.3 nm) of the Lpa-GGH molecule in its extended conformation on Au (see Modeling section). After exposure to $Cu^{2+}$, the thickness is not significantly modified 2.2 ± 0.2 nm. These values are consistent with formation of a highly dense SAM on the $^{TS}$Au surface. The low-density peptide (LDP) monolayers have a thickness of 1.8 ± 0.2 nm indicating an average tilt angle of ~40° to the surface normal. After exposure to $Cu^{2+}$, the thickness is not significantly modified 1.7 ± 0.2 nm. We refer, in the following of the text, to the two SAMs as HDP and LDP, respectively.

XPS analysis of the HDP SAMs. We observed $Au_{4f\ 7/2}$ and $Au_{4f\ 5/2}$ (80.0 and 87.7 eV, respectively), $C_{1s}$ (complex signal containing 3 peaks at 284.6, 286.4 and 288.3 eV), $N_{1s}$ (400.2 eV), $O_{1s}$ (531.7 eV). Figure S2 shows $S_{2p3/2}$ and $S_{2p1/2}$ at 161.9 and 163.1 eV, respectively (exclusively sulfur atoms bound to Au) indicating that dithiolane linkers are fully chemisorbed on gold through dithiolate bonds. The ratios of measured atomic concentrations were in agreement with the expected values (table S1). After $Cu^{2+}$ exposure, the same elements are present and the $Cu_{2p3/2}$ and $Cu_{2p1/2}$ peaks (932.6 and 952.4 eV, respectively) are clearly observed (figure S3). These peaks are significantly shifted compare to typical XPS $Cu^{2+}$ peak (933.6 eV), which is attributed to copper chelation. These results confirm the formation of the Lpa-GGH SAM and its complexation with $Cu^{2+}$ They are in good agreement with our previous report.[14]

Tapping mode-AFM images of the bare $^{TS}$Au surface, of the HDP SAMs before and after the $Cu^{2+}$ exposure were recorded during the KPFM measurements (see Methods). These images (Figure 2) reveal a rather homogeneous SAM formation and no modification after the $Cu^{2+}$ exposure. The root mean square (rms) of roughness is a parameter to quantify this homogeneous SAM formation. We have measured the same rms roughness for the $^{TS}$Au, the HDP SAM (0.40 ± 0.02 nm) and the HDP+$Cu^{2+}$ SAM (0.39 nm ± 0.01 nm). These results again confirm the formation of highly dense Lpa-GGH SAMs.

**Electrical characterization of the GGH SAMs**

For the electrical measurements at the nanoscale, we used two instruments in parallel, one for the C-AFM and one for the KPFM measurements (see Methods). The objective was to avoid any variations in the set-up experiments during the complete sequence of measurements. Since the tip for the C-AFM and KPFM measurements were not the same, using the same instrument and changing the tip back and forth makes difficult to recover the same parameters. The parameters of these two instruments were fixed and kept constant for the complete set of measurements (to ensure a reliable comparison between reference sample ($^{TS}$Au-Lpa-GHH SAMs before $Cu^{2+}$ exposure) and after the $Cu^{2+}$ exposure for both protocols A and B. Thus, measurements of a given set of samples were done in the same session, same configuration, conditions, and with the same tip.

We measured the work function modification of the HDP SAM upon $Cu^{2+}$ exposure by macroscopic Kelvin probe (KP, see Methods) and at the nanoscale by Kelvin probe force microscopy (KPFM). The contact potential difference (CPD) is reduced by about 70 meV (considering the mean of the CPD distribution, see figure S4) after the $Cu^{2+}$ exposure, which corresponds to an increase of the WF by 70 meV (Fig. 3). The macroscopic KP measurements give an increase of WF by 75 meV after $Cu^{2+}$ exposure of HDP surface (Fig. 3). KPFM measurements for the LDP SAM shows an increase of the WF by about 30 meV (CPD values in Figure S4, Supporting Information) in agreement with the trend observed for the denser SAM, albeit the WF increase is smaller (see Fig. 3). The same trends were observed for the control experiments (protocol B) for both LDP and HDP SAMs (figure S4 in the Supporting Information). The change in WF for peptide layer after exposure to $Cu^{2+}$ is explained by peptide-copper complexation, which is attributed to peptide conformational changes. These conformational changes alter the dipole of the layer and hence have a significant influence on the WF. The KP and KPFM similar values proves that the microscopic and macroscopic KP analysis are in agreement thus indicates the homogeneity of the layer and supports effect of conformational changes.

The current-voltage (I-V) curve of the HDP SAM and the HDP+$Cu^{2+}$ SAMs were measured by C-AFM and the current histograms at 1 V constructed form 300 I-V measurements are shown in Figure 4. The histograms are fitted by log-normal distributions. The parameters, log-mean current (log μ) and log-standard deviation (log σ) are given in Table 1. The Cu$^{2+}$ exposure induces a slight decrease of log μ (at 1V) from -11.12 (i.e. 7.6x10$^{-12}$ A) to -11.90 (1.26x10$^{-12}$ A). The same trend was observed for the control experiment, see Figure S5. Thus, the Cu$^{2+}$ exposition induces a slight decrease of the current through the HDP SAM. However, in that case of the LDP SAM, the C-AFM measurements reveal an increase of the current after Cu$^{2+}$ exposure (Fig. 5, Table 1), from log μ (at 2V) = -11.77 (i.e. 1.7x10$^{-12}$ A) to -10.77 (1.7x10$^{-11}$ A), with again a similar trend (log μ (at 2V) -10.13 (7.4x10$^{-11}$ A)) for the control experiments (Figure S6).

|  | HDP | | LDP | |
|---|---|---|---|---|
|  | log μ | log σ | log μ | log σ |
| $^{TS}$Au-Lpa-GGH | -11.12 (7.6×10$^{-12}$ A) | 0.7 | -11.77 (1.7×10$^{-12}$ A) | 0.61 |
| $^{TS}$Au-Lpa-GGH+Cu$^{2+}$ | -11.90 (1.26×10$^{-12}$ A) | 0.52 | -10.77 (1.7×10$^{-11}$ A) | 0.75 |

Table 1. Fitted parameters, log-mean current (log μ), and log-standard deviation (log σ) of the normal distributions shown in Figs. 4, 5, S5 and S6 (Supporting Information).

**Modeling**

We have first addressed for the sake of reference the electronic structure and the corresponding work function changes for the case studied in a previous publication,[14] where only structural issues were discussed. As we showed, Lpa-GGH covalently bonds to the Au(111) surface by breaking the di-sulfide bond and, upon $Cu^{2+}$ chelation, the molecule experiences a conformational change from a nearly linear configuration to a square-planar coordination where four N atoms englobe the $Cu^{2+}$ ion. In the different panels of Fig. 6 various components of the projected electronic density of states (PDOS) are displayed for both geometrical conformations of the Lpa-GGH without and with $Cu^{2+}$ binding. Shown are projections on the Au substrate (grey background), on the Lpa-GGH molecule, and on the Cu and N atoms belonging to the molecule. The states close to the Fermi energy EF have a strong contribution from the N atoms and, upon chelation of $Cu^{2+}$, also Cu-derived states emerge there (see the lower panel of Fig. 6). As a result, there is considerable hybridization of N-based and Cu-based orbitals. The hybridization takes place mostly between the N p-states and Cu d-states, with some additional weight coming from the O p-states (As shown in Figure S7 of the Supporting Information). We also remark that the state slightly above the Fermi level has a small electronic occupation and may be interpreted as a metal (Cu)-induced gap state, where gap refers to the molecular HOMO-LUMO gap of the isolated GGH molecule. We may expect that this state should contribute to the electronic transport in the linear response (low-bias) limit, thus providing a transport channel that would be absent otherwise. Overall, the $Cu^{2+}$ binding shifts the states below the Fermi level closer to it.

To shed further light into the experimental findings related to SAMs with low and high packing density, we have also considered different scenarios for the $Cu^{2+}$ ion binding in the Lpa-GGH/Au(111) system; they are labeled as conformations I-IV in Fig. 7. They differ from each other in the number of N atoms (1 to 4) involved in the binding to $Cu^{2+}$, conformation C-IV corresponding to the tetradentate N-donor square planar complex previously discussed.[14] The other complexes, C-I (monodentated), C-II (bidentated), and C-III (tridentated) turned out to be energetically less favorable when considering the Cu-binding to single GGH molecules. The binding process within an HDP monolayer may favor such complexes because of steric hindrance that prevents the formation of the more stable square planar complex C-IV. The corresponding work functions (WF) of these systems have been calculated by using the relation WF = -e F($\infty$) – $E_F$, where $E_F$ is the bulk Fermi energy, e is the electron charge, and F($\infty$) is the asymptotic electrostatic potential in vacuum. F($\infty$) is extracted from the in-plane-averaged potential $\langle \Phi(z) \rangle = (1/A) \iint_A dx\,dy\, \Phi(x,y,z)$ with A being the area of the surface unit cell. We have found that the WF of the bare Au(111) surface (WF=5.32 eV) decreases after attachment of Lpa-GGH to a value of 3.98 eV (see the inset in Fig. 7). Upon ion binding, the WF increases again smoothly from Case IV to Case I, attaining in the latter case its largest value of ~5.62 eV, i.e. situations with a HDP SAM may yield larger WFs as for LDP.

The change in the work function after ion binding obviously correlates with the modification of the interface dipole $\Delta\mu$, as follows from Helmholtz equation: $\Delta WF = (e/\varepsilon_0 A)\Delta\mu$. In the most stable situation, case IV, we found a strong modification of the interface dipole, manifesting in a change from +7.3 Debye for Lpa-GGH on Au(111) to -1.3 Debye after binding of $Cu^{2+}$. The molecular dipole moment μ is computed according to the relation $V(+\infty) - V(-\infty) = \frac{e\mu}{\varepsilon_0 A}$, where V(+∞) and V(-∞) are the asymptotic electrostatic potentials on both sides of the SAM, e is the electron charge, A is the surface area, and $\varepsilon_0$ the vacuum dielectric constant. These are easily obtained from our electronic structure calculations, since the electrostatic potential reaches its asymptotic values within a distance of few Å from the metal-molecule interface.[23] These changes of the dipole moment relate to the charge rearrangement at the molecule/surface system resulting from ion binding and the associated conformational change. The relatively strong change in the WFs of roughly 1 eV after ion binding found in our calculations may be related to the fact that no depolarization effects were taken into account in our calculations. In Figures S8 and S9 (Supporting Information) the PDOS for all cases are shown, projected on individual atomic subsets as well as orbital-resolved. Here, we see that very close to the Fermi energy states arising from Cu(d)-N(p) state hybridization are dominating. In configuration IV, there is additionally a split state slightly above the Fermi energy, which contains contributions from N p-, O p-, and Cu p-states (see panel d) in Figure S8).

**DISCUSSION**

The increase of the WF after $Cu^{2+}$ exposition measured by KP and KPFM are in good agreement (around 70 meV, see Fig. 3) and qualitatively supported by theory (increase by 1 eV if we consider $Cu^{2+}$ ions surrounded by a ring with 4 N atoms, Fig. 7). The experimental values are lower than theory, which can be explained by the fact that the depolarization effect[24-25] in the SAM, which is known to reduce the dipole per molecule,[26-28] were not taken into account in the calculation (single molecule).

For the current-voltage measurements by C-AFM, we first consider that after $Cu^{2+}$ exposure, the Lpa-GGH peptide undergoes a conformational change forming tetradentate N-donor square planar complex (C-IV complex) of the $Cu^{2+}$ ion as reported in a previous work.[14] According to the DFT calculations of C-IV (Fig. 6), the HOMO moves closer (at ~ -0.55 eV) to the metal Fermi energy $E_F$, and an additional, partially occupied state appears just above $E_F$ (at ~ +0.26 eV). Consequently, we would expect an increase of the current through the molecular junction, since these additional molecular states can fall in the energy window opened by applying a voltage between the two electrodes of the $^{TS}$Au/SAM/C-AFM tip junction. This scenario is consistent with the C-AFM measurements on the less dense SAM (Fig. 5). We note that this effect (increase of electron transport) resembles that recently observed when "doping" a polypeptide (7-alanine) by substituting with a single tryptophan unit introduces an energy level near the Fermi energy.[4] This is also consistent with the observation that incorporating chelated metal atoms in a molecular chain significantly improves the conductance of the molecular junction.[29]

We have observed (Fig. 4) only a slight decrease of the current for the denser SAM. We explain this feature by assuming that, in the present case, the steric hindrance prevents a complete folding and only a partial coordination (i.e. cases I to III depicted in Fig. 7) is allowed. In that case, the DFT calculations (see panels a-c in both Figs. S6 and S7 in the Supporting Information) show that no additional state appears above the Fermi energy, and we do not expect a clear increase of the current in the molecular junction. However, even in these cases, the electronic structures calculated for the $Cu^{2+}$ induced partial conformational changes (forming C-I, C-II, and C-III complexes) show that the HOMO level is moving closer to the Au Fermi energy as in the C-IV complex case (Figs. S6 and S7 in the supporting information). In principle, this feature alone can also induce an increase of the current through the molecular junction, which is not observed for the denser sample (Fig. 4). This point deserves more detailed calculations of the electron transfer probability and I-V curves taken into account the molecule/ tip (PtIr) interface (for example, the contact with high WF tip can counterbalance the shift of the HOMO level due to charge transfer at the molecule/tip interface), while the presence of an additional electron transmission channel (as in C-IV) induces the observed increase of current.

Finally, we assume that the surface roughness determines the ability of the peptide to acquire a full conformational change to form C-IV complex (square planar coordination). In this study we used ultra flat gold while in our previous study[14] a gold substrate with rougher morphology was used. We assumed that a denser monolayer could be assembled on flat $^{TS}$Au substrate (rms roughness of 0.4 nm, Fig. 2) while the rougher surface (evaporated Au surface with rms roughness of ~0.9 nm) leads to a less packed monolayer. The lower density GGH monolayer allows the full conformational change to yield C-IV complex while this complex is less likely to form in the high-density GGH monolayer on $^{TS}$Au substrate. Our TM-AFM (Fig. 2) and ellipsometry measurements show that there is no significant morphological or topographic changes induced by $Cu^{2+}$ binding thus suggests that the C-IV complex can't be formed on the high-density GGH monolayer.

**CONCLUSION**

The LDP assembled on gold results in an increase of work function and enhanced conductivity in response to $Cu^{2+}$ complexation. On the other hand, the HDP assembled on gold results in larger increase in work function and decrease in electrical current passing through this monolayer. We attribute these differences to the ability of the peptide to adopt a square planar coordination with copper in the LDP and the steric hindrance in the HDP that prevents such conformational changes. Our DFT calculations proved that the significant increase of the current through the LDP-Cu monolayer can result from the movement of the molecular orbitals closer to the gold Fermi energy of the electrodes and an additional electronic state that appears above the gold Fermi energy that is unique to the C-IV complex.

ASSOCIATED CONTENT

**Supporting Information**. The following information is available in the Supporting Information : protocol of measurements, XPS data, CPD values from KPFM measurements, current-voltages curves and hisotgrams for the HDP and LDP control SAMs (protocol B), additional PDOS calculated for the Lpa-GGH and Lpa-GGH SAMs with different conformations of the chelated $Cu^{2+}$ ion.

AUTHOR INFORMATION

**Corresponding authors.**

dominique.vuillaume@iemn.fr

rafael.gutierrez@tu-dresden.de

shlomo.yitzchaik@mail.huji.ac.il

**ORCID**

Evgeniy Mervinetsky : 0000-0002-4373-1263

Israel Alshanski : 0000-0002-9310-1921

Stephane Lenfant : 0000-0002-6857-8752

Gianaurelio Cuniberti : 0000-0002-6574-7848

Mattan Hurevich : 0000-0002-1038-8104

Shlomo Yitzchaik : 0000-0001-5021-5139

Dominique Vuillaume : 0000-0002-3362-1669

**Notes**

The authors declare no competing financial interest.

**Acknowledgements**

The authors would like to thank RECORD-IT project. This project has received funding from the European Union's Horizon 2020 research and innovation program under grant agreement No 664786. SY is the Binjamin H. Birstein Chair in Chemistry. This work has also been partly supported by the German Research Foundation (DFG) within the Cluster of Excellence "Center for Advancing Electronics Dresden". We acknowledge the Center for Information Services and High Performance Computing (ZIH) at TU Dresden for computational resources.

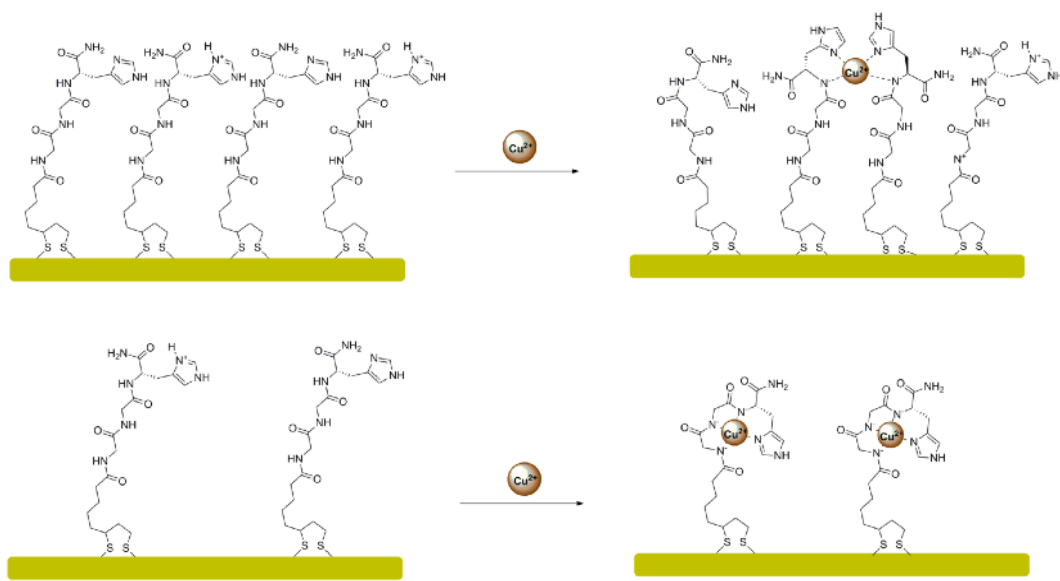

Figure 1. Schematic representation of conformational changes of the Lpa-GGH monolayers in response to $Cu^{2+}$ chelation. Top: sterically hindered chelation for high-density peptide (HDP) monolayer; Bottom: unperturbed chelation for lower density peptide (LDP) monolayer.

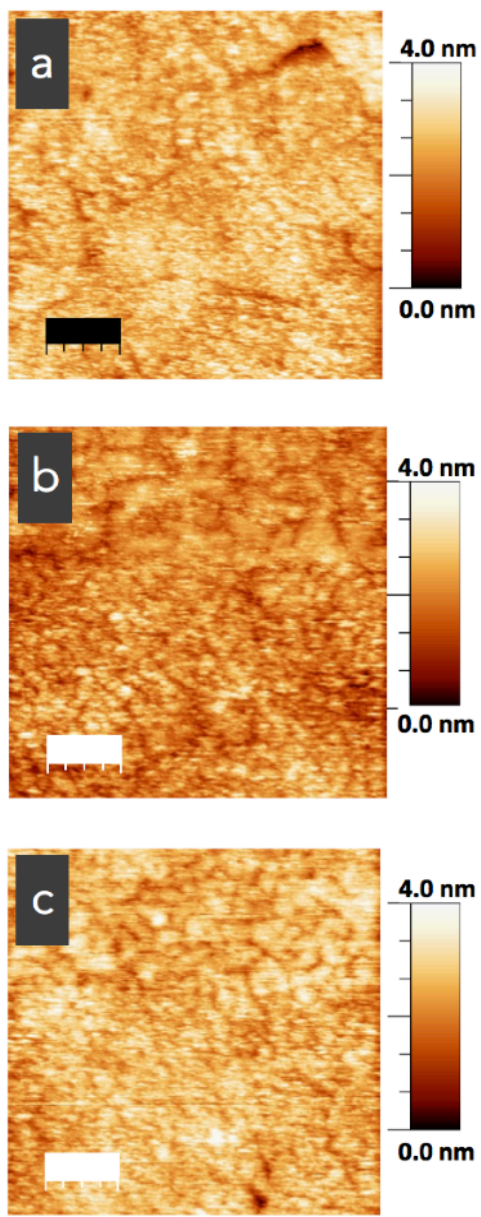

Figure 2. Tapping Mode-AFM images of (a) the naked $^{TS}$Au substrate, rms roughness of 0.40 ±0.02 nm, (b) the HDP SAM, rms roughness of 0.40 ±0.02 nm,

and (c) HDP SAM after $Cu^{2+}$ exposure, rms roughness of 0.39 ±0.01 nm. Scale bars are 200 nm.

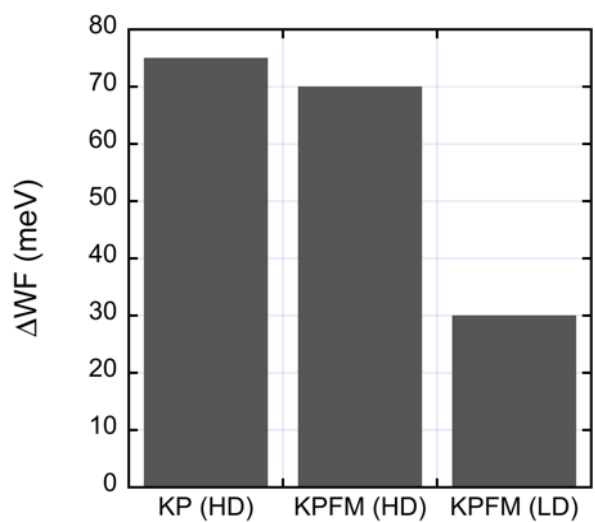

Figure 3. Work function variations (average values) for the HDP SAM exposed to $Cu^{2+}$ measured by KP and KPFM and for LDP SAM measured by KPFM.

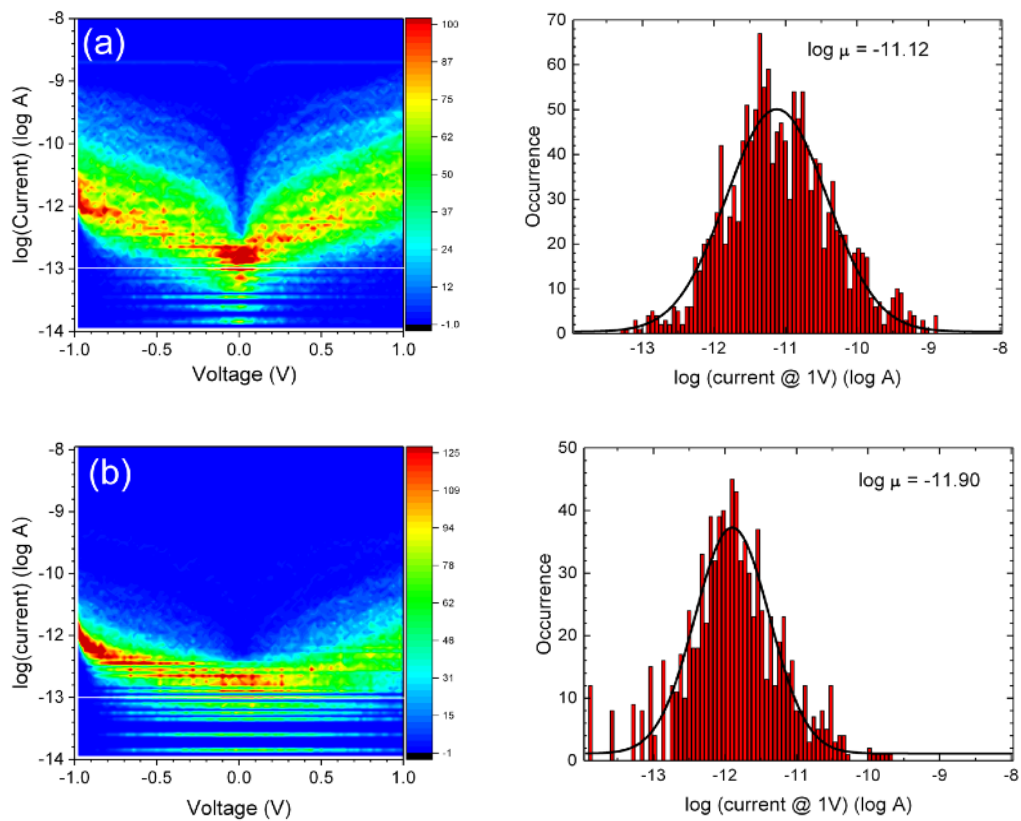

Figure 4. 2D histograms of the current-voltage curves (300 I-V traces, tip loading force 50 nN) and corresponding current histograms at 1V for: (a) HDP SAM, (b) HDP SAM exposed to $Cu^{2+}$. The current histograms at 1 V are fitted (black lines) with log-normal distributions (fitted parameters in Table 1). In the I-V histograms, the white lines are the sensitivity limit of the C-AFM system (~0.1 pA).

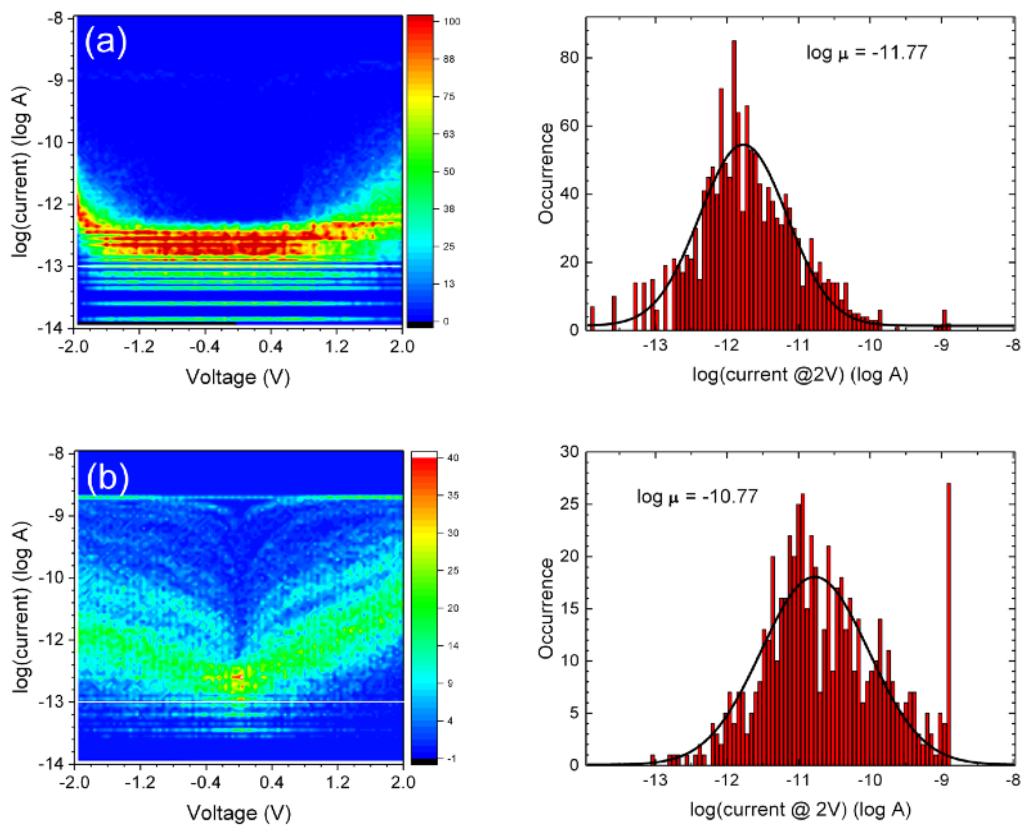

Figure 5. 2D histograms of the current-voltage curves (300 I-V traces, tip loading force 29 nN) and corresponding current histograms at 2V for the less densely packed SAM (at 10 μM): (a) LDP SAM, (b) LDP SAM exposed to $Cu^{2+}$. The current histograms at 2 V are fitted (black lines) with log-normal distributions (fitted parameters in Table 1). In the I-V histograms, the white lines are the sensitivity limit of the C-AFM system (~0.1 pA).

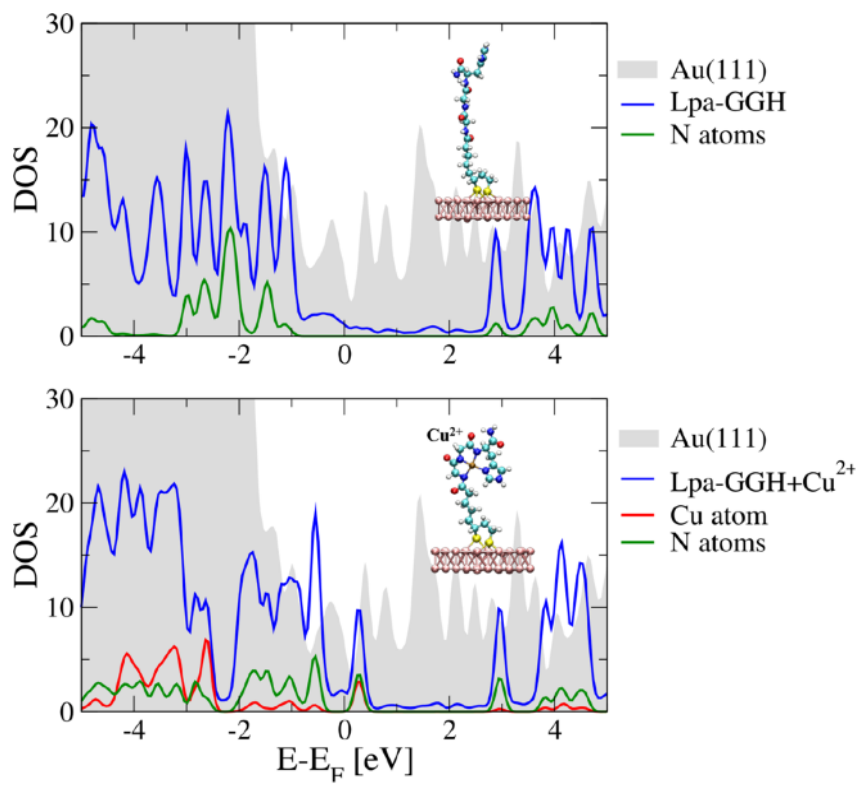

Figure 6. Projected Density of States (PDOS) on various atomic subsets for the Lpa-GGH molecule chemisorbed on Au(111) (upper panel) and upon binding a $Cu^{2+}$ ion in the C-IV complex (lower panel). In this later case, electronic states shortly above the Fermi energy (set at zero) result from the hybridization between Cu d-states and N p-states (see also Figure S7 in Supporting Information). The corresponding structural conformations are inserted in both panels.

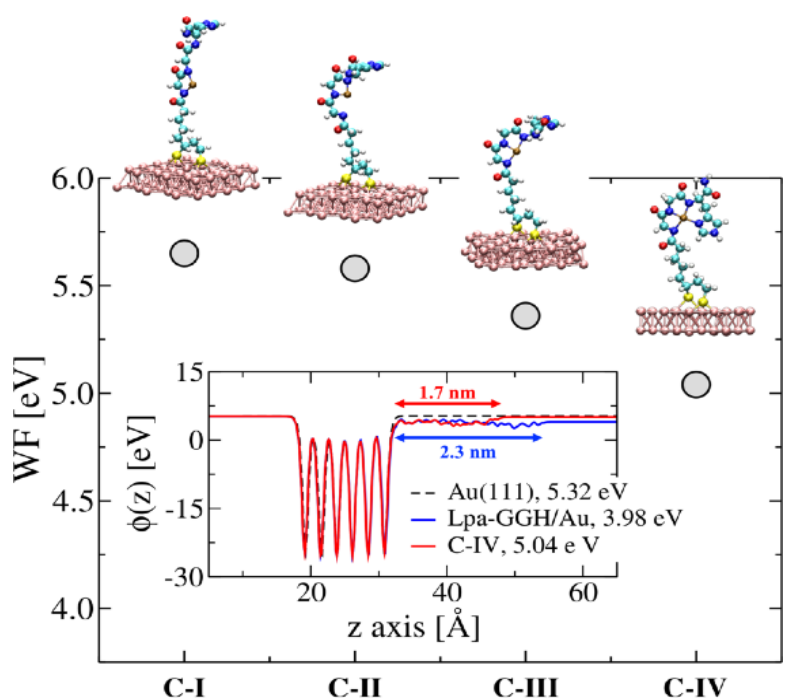

Figure 7. Calculated work functions for the four different conformations considered in this study C-I to C-IV (Cu coordinated to 1, 2, 3 and 4 N atoms, respectively) and described in the main text. Independently of the studied conformation, the WF always increases upon $Cu^{2+}$ chelation. Inset: Variation of the electrostatic potential along the z-direction for the different relevant cases: bare Au(111), Lpa-GGH on Au(111), and configuration C-IV: Lpa-GGH/$Cu^{2+}$ on Au(111). Indicated are also the computed values of the work functions for the three cases. After $Cu^{2+}$ chelation the WF increases by 1.06 eV for the most stable configuration (C-IV). The horizontal arrows indicate the change in molecular length related to the molecular conformational change (going from 2.3 nm down to 1.7 nm).

7. Scullion, L.; Doneux, T.; Bouffier, L.; Fernig, D. G.; Higgins, S. J.; Bethell, D.; Nichols, R. J. Large Conductance Changes in Peptide Single Molecule Junctions Controlled by Ph. *J. Phys. Chem. C* **2011**, *115*, 8361-8368.

8. Xiao, X.; Xu, B.; Tao, N. Conductance Titration of Single Peptide Molecules. *J Am Chem Soc* **2004**, *126*, 5370-5371.

9. Uji, H.; Morita, T.; Kimura, S. Molecular Direction Dependence of Single-Molecule Conductance of a Helical Peptide in Molecular Junction. *Phys. Chem. Chem. Phys.* **2013**, *15*, 757-760.

10. Kitagawa, K.; Morita, T.; Kimura, S. Electron Transfer in Metal-Molecule-Metal Junction Composed of Self-Assembled Monolayers of Helical Peptides Carrying Redox-Active Ferrocene Units. *Langmuir* **2005**, *21*, 10624-10631.

11. Kitagawa, K.; Morita, T.; Kimura, S. Molecular Rectification of a Helical Peptide with a Redox Group in the Metal-Molecule-Metal Junction. *The journal of physical chemistry B* **2005**, *109*, 13906-13911.

12. Devillers, C. H.; Boturyn, D.; Bucher, C.; Dumy, P.; Labbé, P.; Moutet, J.-C.; Royal, G.; Saint-Aman, E. Redox-Active Biomolecular Architectures and Self-Assembled Monolayers Based on a Cyclodecapeptide Regioselectively Addressable Functional Template. *Langmuir* **2006**, *22*, 8134-8143.
32

**TOC graphics**

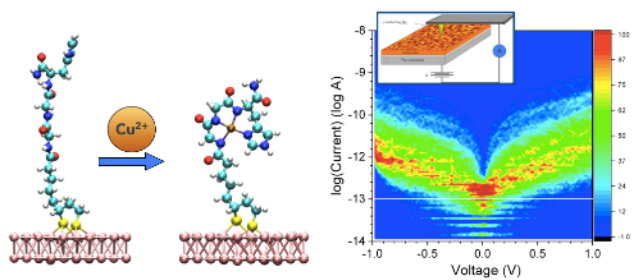

# Electron transport through tripeptides self-assembled monolayers

Evgeniy Mervinetsky,[1,2] Israel Alshanski,[1,2] Stephane Lenfant,[3] David Guerin,[3] Leonardo Medrano Sandonas,[4,6] Arezoo Dianat,[4] Rafael Gutierrez,[4] Gianaurelio Cuniberti,[4,5,6] Mattan Hurevich,[1,2] Shlomo Yitzchaik[1,2,] & Dominique Vuillaume[3]

1) Institute of Chemistry, The Hebrew University of Jerusalem, Safra campus, Givat Ram, Jerusalem, 91904, Israel.
2) Center for Nanoscience and Nanotechnology, The Hebrew University of Jerusalem, Safra campus, Jerusalem, 91904, Israel.
3) Institute for Electronics Microelectronics and Nanotechnology, CNRS, Univ. Lille, 59652, Villeneuve d'Ascq, France.
4) Institute for Materials Science and Max Bergmann Center of Biomaterials, TU Dresden, 01069 Dresden, Germany.
5) Dresden Center for Computational Materials Science, TU Dresden, 01062 Dresden, Germany.
6) Center for Advancing Electronics Dresden, TU Dresden, 01062 Dresden, Germany

# Supporting information

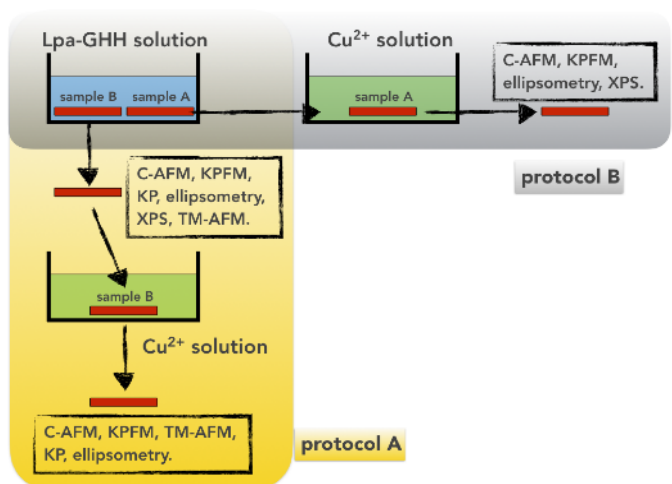

**Figure S1**. Schematic description of the sample preparation and measurement protocols.

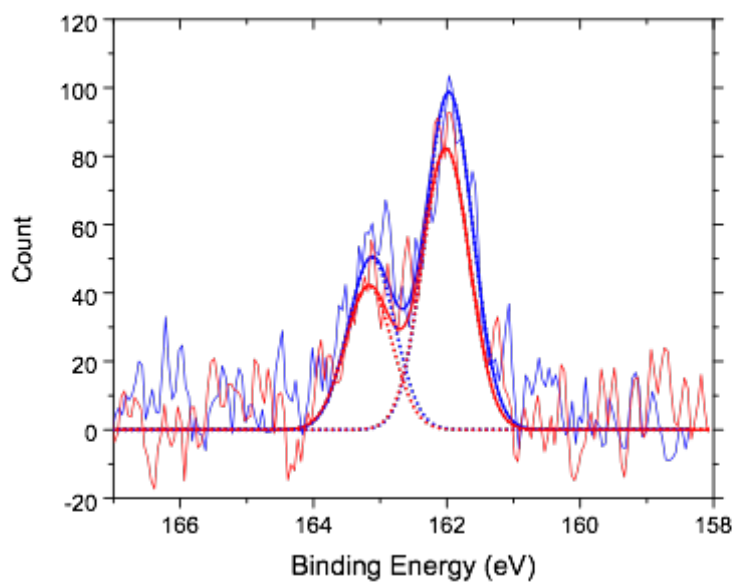

**Figure S2**. The $S_{2p}$ signal by XPS of HDP SAM (blue) before and (red) after $Cu^{2+}$ exposure exhibited only sulfur atoms bound to gold ($S_{2p3/2}$ at 161.9 eV and $S_{2p1/2}$ at 163.1 eV) indicating that molecules are fully chemisorbed on gold.

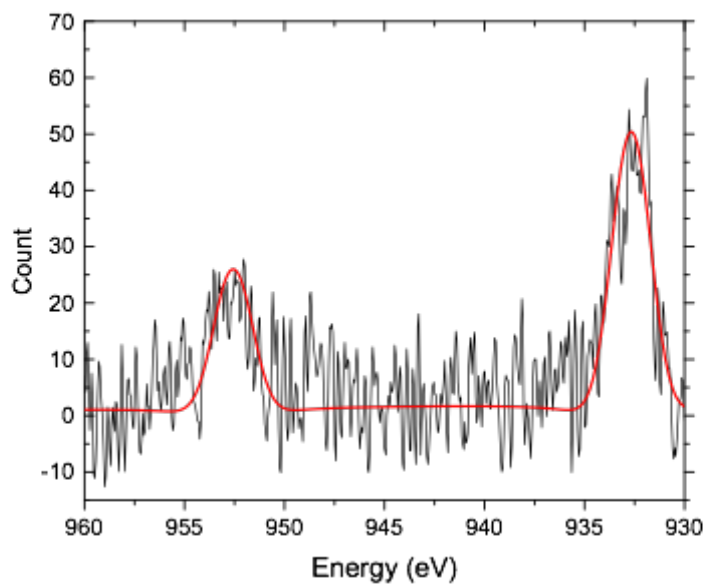

**Figure S3.** The $Cu_{2p}$ signal ($Cu_{2p1/2}$ at 952.4 eV and $Cu_{2p3/2}$ at 932.6 eV), shifted compare to free $Cu^{2+}$ (933.6 eV), was only observed in the SAM exposed to $Cu^{2+}$.

|  | **HDP** | | **HDP + $Cu^{2+}$** | |
|---|---|---|---|---|
|  | *BE (eV)* | *$A_{corrected}$* | *BE(eV)* | *$A_{corrected}$* |
| *C1s* | 284.6, 286.4, 288.3 | 4023 | 284.6, 286.4, 288.3 | 3665 |
| *N1s* | 400.2 | 1451 | 400.2 | 1276 |
| *O1s* | 531.7 | 1125 | 531.6 | 1209 |
| *$S2p_{3/2-1/2}$* | 161.9, 163.1 | 319 | 161.9, 163.1 | 277 |
| *$Cu2p_{3/2-1/2}$* | - | - | 932.6, 952.4 | 43 |
| N/S = 4.5 (3) <br> N/O = 1.3 (1.5) <br> O/S = 3.5 (2) <br> C/N = 2.7 (3) <br> C/O = 3.6 (4.5) | | | N/S = 4.6 (3) <br> N/O = 1.0 (1.5) <br> O/S = 4.3 (2) <br> C/N = 2.9 (3) <br> C/O = 3.0 (4.5) | |

**Table S1**. Binding energy and concentration of elements by XPS in HDP SAM before and after $Cu^{2+}$ exposure. Corrected area corresponds to peak area divided by the atomic sensitivity factor. Atomic ratios measured by XPS are compared with theoretical values in brackets.

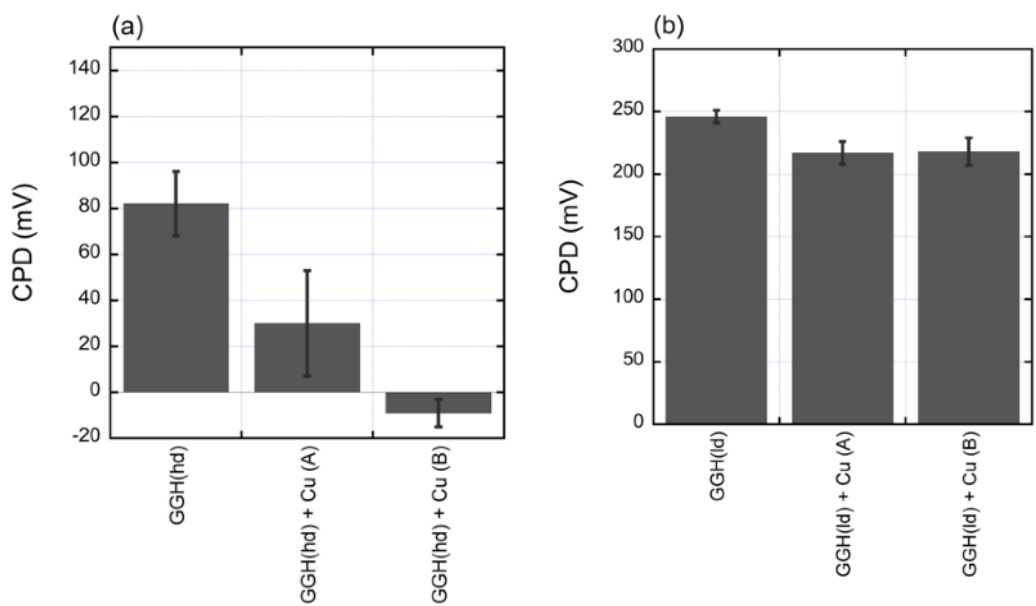

**Figure S4. (a)** CPD measured by KPFM on HDP, HDP + $Cu^{2+}$ protocol A, and HDP + $Cu^{2+}$ protocol B. All values are averaged from 4 KPFM images. We deduce an increase of WF by 52±37 meV (protocol A) and 91±20 meV (protocol B). The average WF increase, reported in Fig. 3 (main text) is 71.5 mV **(b)** CPD measured by KPFM on LDP, LDP + $Cu^{2+}$ by protocol A and LDP + $Cu^{2+}$ protocol B. All values are averaged from 3 KPFM images. We deduce an increase of WF by 29±14 meV (protocol A) and 28±16 meV (protocol B)

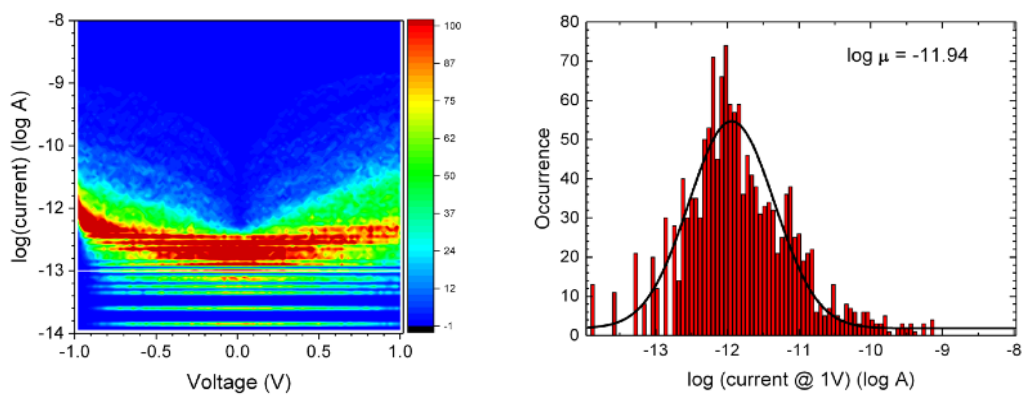

**Figure S5.** 2D histograms of the current-voltage curves (300 I-V traces, tip loading force 50 nN) and corresponding current histograms at 1V for HDP SAM exposed to $Cu^{2+}$ (protocol B). The current histograms at 1 V are fitted (black lines) with log-normal distributions (fitted parameters in Table 1). In the I-V histograms, the white line is the sensitivity limit of the C-AFM system (~0.1 pA).

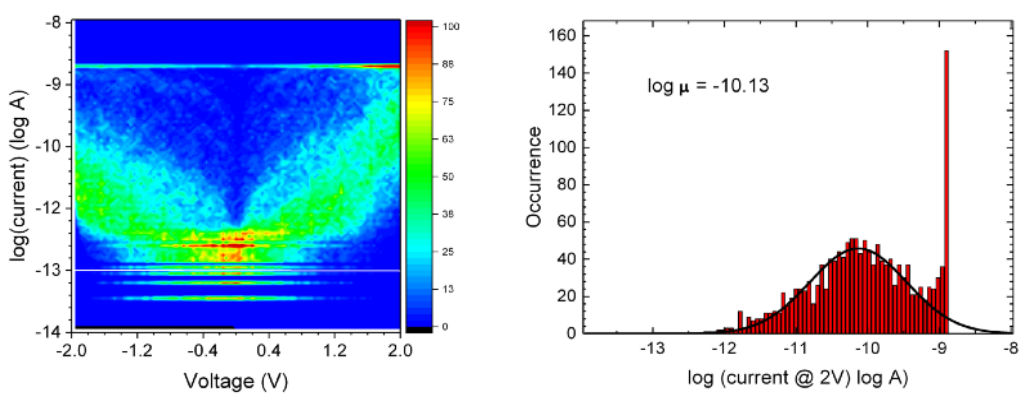

**Figure S6.** 2D histograms of the current-voltage curves (300 I-V traces, tip loading force 29 nN) and corresponding current histograms at 2V for LDP SAM exposed to $Cu^{2+}$ (protocol B). The current histograms at 1 V are fitted (black lines) with log-normal distributions (fitted parameters in Table 1). In the I-V histograms, the white lines is the sensitivity limit of the C-AFM system (~0.1 pA)

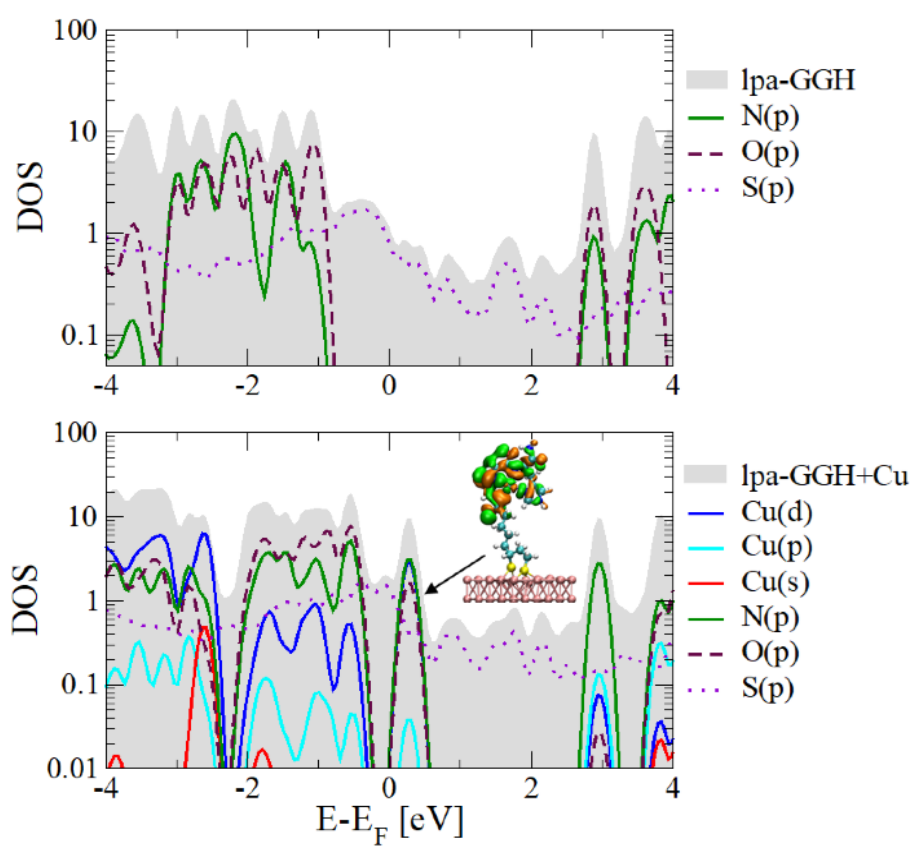

**Figure S7.** Orbital resolved projected density of states for Lpa-GGH/Au(111) before (upper panel) and after (lower panel) $Cu^{2+}$ ion binding for the C-IV complex (see also the main text). The inset in the lower panel shows the spatial charge density distribution of the state closest to the Fermi energy, which is derived from the hybridization of Cu d- and N p-states with some contribution from O p-states, too.

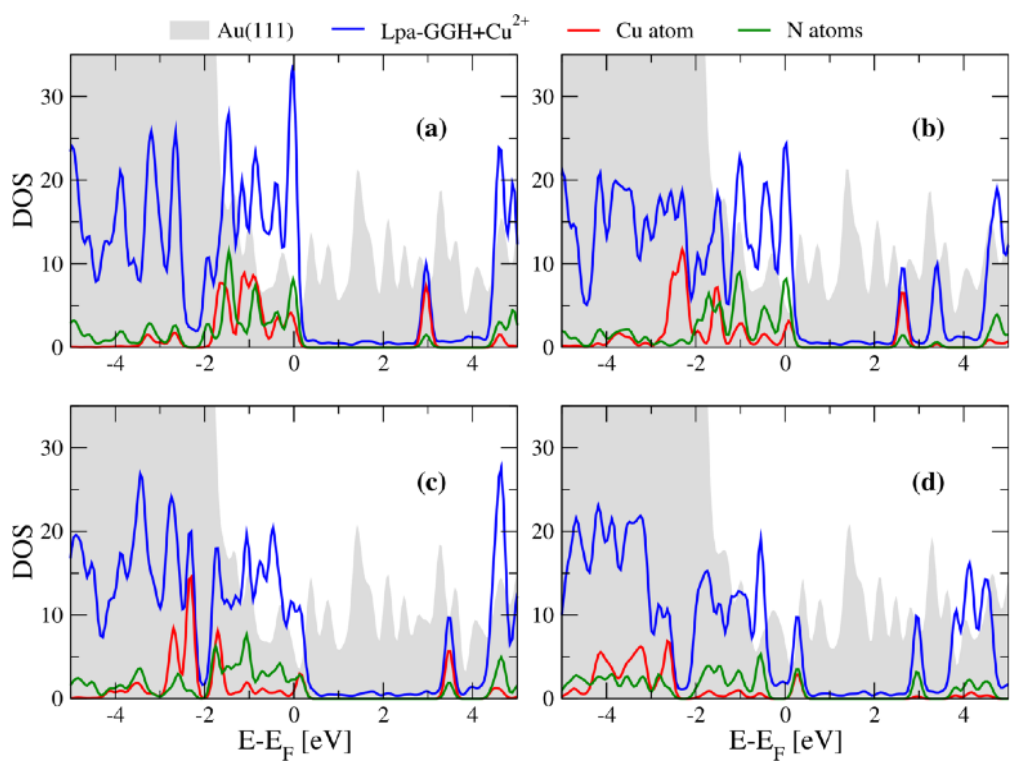

**Figure S8.** Projected density of states for the four possible configurations studied in the main text, in which $Cu^{2+}$ can bind to Lpa-GGH/Au(111). Panels a-d correspond to the conformations shown in Fig. 7 of the main text (a) C-I, (b) C-II, (c) C-III and (d) C-IV. We show the DOS projected on the Lpa-GGH+$Cu^{2+}$ complex as well as on the Cu atom and all the N atoms of the molecule. With increasing coordination of the Cu to the N atoms in the ring region, the spectral contribution at the Fermi energy is continuously reduced from a to d. Notice that in the latter case almost no Cu and N weight is observed at the Fermi energy.

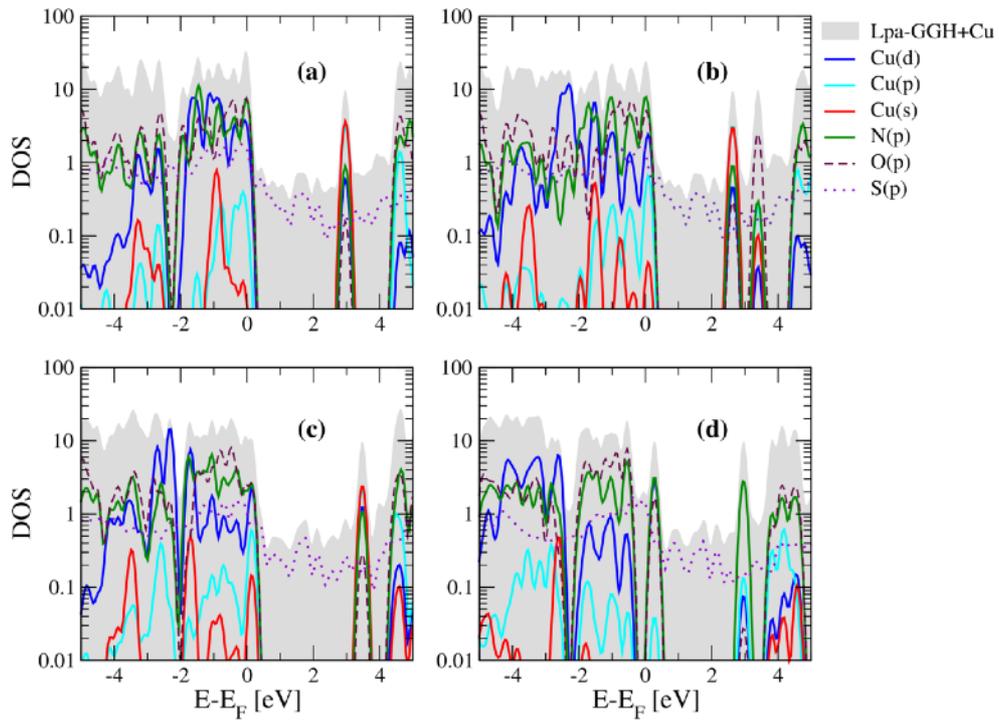

**Figure S9.** Orbital resolved projected density of states (PDOS) of the possible configurations a-d (see Fig. 7 in the main text) for the study of $Cu^{2+}$ ion location in the Lpa-GGH/Au(111) system. Shown are the contributions from Cu s-, p- and d-states, and from the p-states by O, N, and S.